\documentclass[showpacs,prb,superscriptaddress,twocolumn,floatfix]{revtex4}
\usepackage{graphicx,float,amsmath}

\begin{document}

\title{Quantum transport through carbon nanotubes: proximity induced and intrinsic superconductivity.}

\author{ A. Kasumov$^{1,2,*}$,  M. Kociak$^1$, M. Ferrier $^1$, R. Deblock$^1$,  S.~Gu\'eron$^1$, B. Reulet$^1$, I. Khodos$^2$, O. St\'ephan $^1$, and H. Bouchiat$^1$  \\}
\address{Laboratoire de Physique des Solides, Associ\'e au CNRS,
  B\^atiment 510, Universit\'e Paris-Sud, 91405, Orsay, France.\\
  $^2$Institute of Microelectronics Technology and High Purity Materials,Russian Academy of Sciences, Chernogolovka 142432 Moscow Region, Russia.\\
  $*$Present address: RIKEN, Hirosawa 2-1, Wako, Saitama, 351-0198 Japan}

\begin{abstract} We report
 low temperature transport measurements on  suspended  single walled carbon nanotubes (both individual tubes and ropes).  The technique we have developed, where tubes are  soldered on low resistive metallic  contacts across a slit, enables a good characterization  of the samples by transmission electron microscopy.  It is possible to obtain individual tubes with a room temperature resistance smaller than $40$ k$\Omega$, which remain metallic down to very low temperatures.
 %with only $20 \% $ increase of resistance between room temperature and 1K%. 
When the contact pads are superconducting, nanotubes exhibit  proximity induced superconductivity with surprisingly large values of supercurrent. We have also recently observed  intrinsic superconductivity in ropes of single walled carbon nanotubes connected to normal contacts, when the distance between the normal electrodes is large enough, since otherwise superconductivity is  destroyed by (inverse) proximity effect. These experiments indicate the presence of attractive interactions in carbon nanotubes which overcome Coulomb repulsive interactions at low temperature, and enables investigation of superconductivity in a  1D limit never explored before.
Pacs 74.10.+v, 74.70.-b, 74.50.+r

\end{abstract}

\maketitle

\section{Introduction}
The hope to use molecules as the ultimate building blocks of electronic circuits motivates the quest to understand electronic transport in thinner and thinner wires, ideally with one or two  conduction modes.  However a number of physical phenomena tend to drive such one-dimensional (1D) metallic wires to an insulating state at low temperature. The most basic is the great sensitivity of 1D transport to disorder: a 1D wire has a localization length of the order of its elastic mean free path.  It has also been shown that most  molecular 1D conductors  undergo a structural Peirls transition to an insulating state at low temperature \cite{jerome}.  Finally the repulsive Coulomb interactions between electrons, which are very weakly screened in 1D, tend to destabilize the 1D Fermi liquid in flavor of correlated states which are also insulating at low temperature. 

The purpose of this paper is to show that carbon nanotubes, because of their special band  structure, escape such a fate and remain conducting over  lengths greater than one micron down to very low temperature. They do not present any detectable Peirls distortion and are weakly sensitive to disorder. Moreover transport through the nanotubes is quantum coherent, as demonstrated  by the existence of strong supercurrents when connected to superconducting contacts. The observation of intrinsic superconductivity in ropes containing a few tens of tubes is even more surprising 
and indicates the presence of attractive pairing interactions which overcome the strong repulsive interactions.

Single walled carbon nanotubes are constituted by a single graphene plane
wrapped into a cylinder.
The Fermi surface of  graphene  is very particular, it is reduced to six discrete points at the corners of the first Brillouin zone \cite{wall}.  As a result, depending on their diameter and their helicity which determine the boundary conditions of the electronic wave functions around the tube, SWNT can be either semiconducting or metallic \cite{dressel}. When metallic they have just two conducting channels, only  very weakly coupled by electron-electron interactions.  They are thus characterized by a low electronic density together with a high Fermi velocity (nearly as high as in copper) and relatively large mean free paths \cite{white}.   These properties make them much sought-after realizations of 1D conductors, or more precisely of 1D ladders with very small transverse coupling \cite{Egger,Kane}. 
At 1D, electron-electron interactions lead to an exotic  correlated electronic state, the Luttinger liquid (LL)\cite{Luttinger} where collective low energy  plasmon-like excitations give rise  to anomalies in the single particle density of states, and where no long range order exists at finite temperature.

Proof of the validity of LL description with repulsive interactions in SWNT was given by the measurement of a resistance diverging as a power law with temperature down to 10 K \cite{bockrath} with an exponent depending on whether contacts are taken on the end or in the bulk of the tube. From these exponents the LL  parameter $g$  measuring the strength of the interactions was deduced to be $g =0.25\pm0.05$. This value which is much smaller than $1$  corresponds to  dominant repulsive interactions. The extrapolation of this  power law behavior down to very low temperature would indicate an insulating state.  However, these measurements were done on nanotubes connected to the measuring leads through tunnel junctions. Because of Coulomb blockade \cite{Grabert}, the low temperature and voltage  transport regime could  not  be explored. 

We have developed a technique  (described in section II) in which measuring pads are connected through low resistance contact to suspended nanotubes \cite{Kasumov2}. It is  then possible to obtain individual tubes with a resistance at room temperature (RT) no larger than $40~k\Omega$, that increases only slightly at low temperatures (typically a $20 \% $ resistance increase between RT and 1 K) (section III).  The resistance versus voltage ($R V$) curves also exhibit weak logarithmic non-linearities at low temperature. However we find no simple scaling law between the voltage and the temperature-dependence of the differential resistance. This suggests that  several different physical mechanisms could contribute to the very low temperature dependence  of the resistance.

In section IV we show that when the contact pads are superconducting, a supercurrent can flow through SWNTs as well as through ropes of SWNTs \cite{Kasumov}.  These experiments indicate  in principle that quantum coherent transport can take place through carbon nanotubes over micron-long scales.
However the unexpectedly  high measured values of the critical current in individual nanotubes, along with non-linearities in the $IV$ characteristics at voltages much higher than the  superconducting gap of the contacts raise the question of possible intrinsic superconductivity.

Finally  in section V we present measurements on long  ropes of SWNTs connected to normal contacts which  indeed show evidence of intrinsic superconductivity  below 0.5 K provided that the distance between normal electrodes  is large enough.  We discuss the one-dimensional character of the transition and the physical parameters  which govern this transition, such as the presence of normal reservoirs, intertube coupling and disorder.

\section{Sample description and preparation}

\begin{figure}
\includegraphics[width=9 cm]{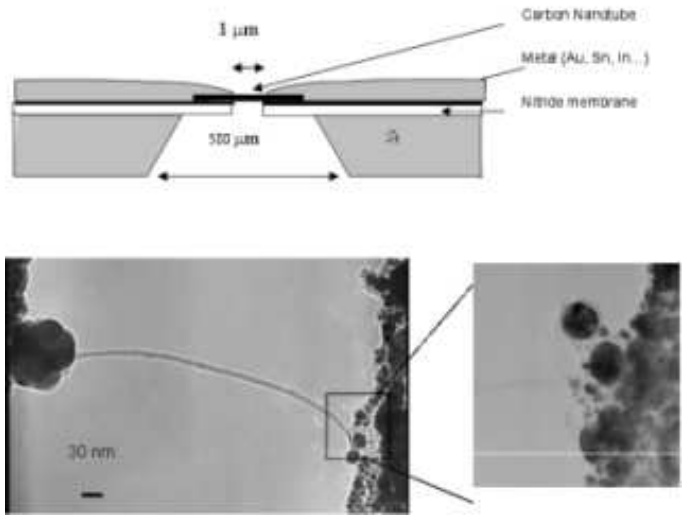}
	\caption{\label{temtube} Principle of the soldering of carbon nanotubes  and schematic cross-section of a device. Below, transmission electronic micrograph (TEM)  of the nanotubes, suspended across a slit between 2 metallic pads; and detailed view of the contact region showing the  metal melted  by the laser pulse.}
\end{figure}

The SWNT are prepared by an electrical arc method with a mixture of 
nickel and yttrium as a catalyst~\cite{Journet,Vaccarini}.
SWNT with diameters of the order of 1.4 nm are obtained. They are 
purified by the cross-flow filtration method~\cite{Vaccarini}.
The tubes usually come assembled in ropes of a few hundred tubes in parallel, but individual tubes can also be obtained after chemical
  treatment with a surfactant~\cite{burghard}. Connection of an individual tube or rope to measuring pads is performed 
according
to the following nano-soldering technique:
A target covered with nanotubes is placed above a 
metal-coated suspended $Si_{3}N_{4}$ membrane,
in which a 0.3~ $\mu m$ by 100~ $\mu m$ slit has been etched with a focused ion beam. 
As described in ref.~\cite{Kasumov2},  we  use a 10 ns-long pulse of a focused UV laser beam (power 
10 kW) to detach a nanotube from the target, which then falls and connects the edges of the slit below. Since the metal electrodes on each side of the slit are locally 
molten, the tube gets soldered into good contacts, and is suspended (see Fig.\ref{temtube}). 

The electrodes are usually bilayers (Re/Au, Ta/Au, Ta/Sn)  or trilayers (Al$_{2}$O$_{3}$/Pt/Au) of non-miscible materials (in liquid or solid phases), in which the lower layer (Re, Ta) has a high melting temperature in order to protect the nitride membrane during the welding of the tube to the top gold layer. Gold was mostly chosen as the solder material because it neither reacts with carbon nor oxidizes. The contact region is pictured in the transmission electron microscope (TEM) view of Fig.\ref{temtube}. Balls of metal molten by the laser pulse can be seen on the edges of the slit.

\begin{table}[t] \centering

\begin{tabular}{|c|c|c|}
\hline
 Contact &composition &$T_c$
\\ \hline

TaAu & 5 nm Ta, 100 nm Au& 0.4 K
\\ \hline
ReAu & 100 nm Re, 100 nm Au& 1 K
\\ \hline
TaSn & 5 nm Ta, 100 nm Sn& 3 K
\\ \hline
CrAu & 5 nm Cr, 100 nm Au& $\le 50~mK$
\\ \hline
PtAu & 5 nm $Al_20_3$, 3 nm Pt, 200 nm Au& $\le 20~mK$
\\  \hline
\end{tabular}
\caption{Main features of the contacts of the investigated nanotubes. $T_c$ denotes their transition temperature when they are superconducting.}
\label{tablecontacts}
\end{table}

\begin{figure}
\includegraphics[width=8cm]{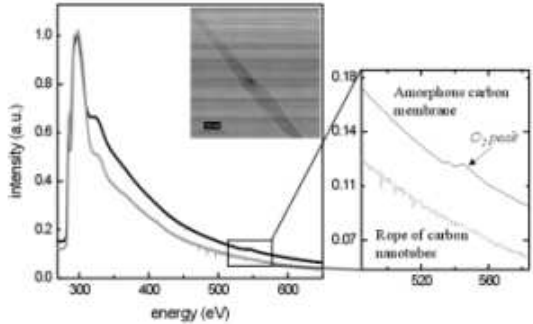}
	\caption{Typical electron energy loss spectrum (EELS) on a carbon nanotube rope (lower curve), compared to the EELS spectrum of an amorphous carbon membrane (upper curve). The peak at 285.5 eV of carbon is visible in both spectra. The oxygen peak (at 540 eV) is much larger in the membrane than on the tubes, and the $O_2$ content on the rope is estimated to be less than $1.5\%$ atomic content.\label{EELS}}
	\end{figure}
	
Because the tubes are suspended, they can be structurally and chemically characterized in a  TEM working at moderate acceleration voltages (100 kV) as shown in Fig.\ref{temtube}.
We measure the length $L$ and diameter $D$ of a rope and estimate the number of tubes $N$ in the rope through $N =(D/(d+e))^2$, where $d$ is the diameter of a single tube ($d$=1.4 nm), and $e$ is the typical distance between tubes in a rope ($e$=0.2 nm). 

We can also check that no metal covers or penetrates the tubes.  In addition we estimate, using electron energy loss spectroscopy (EELS), that chemical dopants such as oxygen are absent to within 2 \% (see Fig. \ref{EELS}). The characterization procedures can of course deteriorate the samples, because of electron irradiation damage  \cite{damage} which can cause local amorphisation and induce an increase in the tubes resistance.

The originality of our fabrication technique lies in the suspended character of the tubes, the quality of the tube-electrode contact, and the ability to characterize the measured sample in a transmission electron microscope.
  
\section{Transport in the normal state}

\subsection{Measurement circuit and sample environment}

The samples were measured in  dilution refrigerators, at temperatures ranging from 300~K to 0.015~K, through filtered lines. Good calibration of the thermometer and thermalisation of the samples are only guaranteed below 1K, so that higher-temperature data are only indicative. Magnetic fields of up to 5 T could be applied perpendicularly to the contacts and the tubes. The resistance was measured by running a small (0.01 nA to 10 nA, 30 Hz) a.c. current though the sample and measuring the corresponding a.c. voltage using lock-in detection.

The two large (roughly $10$ $mm^2$)  metallic contacts on either side of the suspended nanotube constitute, with the ground copper plate on the other side of the silicon wafer, two capacitances of the order 3 pF which help filter-out residual high frequency radiation.

When the tubes were contacted to superconducting electrodes or exhibited intrinsic superconductivity, the investigation of the normal state was done by applying a magnetic field large enough to destroy the superconductivity. 

\subsection{Individual single wall nanotubes}

Depending on their room-temperature resistance, we find that SWNTs can be insulating or metallic at low temperature. The SWNT samples with high (larger than $40$~$
k\Omega$) RT resistance become insulating, with an exponential increase of resistance, as T is decreased. This insulating behavior could be due to the atomic structure of the tube, which would correspond to a semiconducting band structure. It could then lead to a  gap smaller than the expected value at half filling  because of possible doping due by the contacts. It could also be caused by the presence of disorder in a tube (the importance of which is discussed further along). In the following, we focus exclusively on the SWNT samples which remain metallic down to low temperature. They are presented in  table \ref{tableST}. Their resistance at room temperature lies between  $10~k\Omega$ and $70~k\Omega$,  and increases by less than a factor 2 between 300~K and 1~K . 

\begin{figure}
\includegraphics[width=8cm]{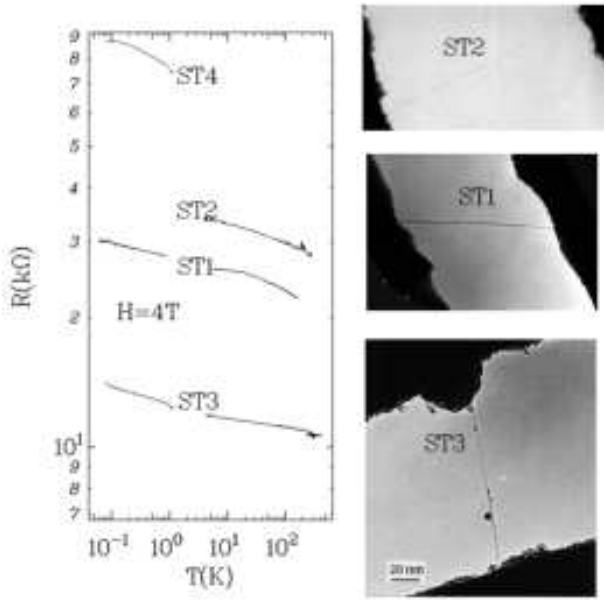}
\caption{\label{RTnorm} Temperature dependence of the resistance of different single tubes: ST1, ST2, and ST4 are mounted on Au/Ta contacts and measured in a magnetic field of 4 T to suppress the superconductivity of the contacts. ST3 is mounted on CrIn contacts. Note the logarithmic scales. TEM micrograph of the corresponding samples. The dark spots are Ni/Y catalyst particles .}
	\end{figure}

The temperature dependence of the zero bias resistance of four samples is shown in Fig. \ref{RTnorm} in a log-log plot.  The  increase of resistance as T is lowered can be described  by a power law with a very small  negative exponent which varies between 0.01  for the less resistive tube and 0.1  for the most resistive one. Note that because of the very small values of these exponents a logarithmic law $R \propto R_0 -\alpha~ log T $ could fit the data just as well. This behavior is observed down to 50 mK without any sign of  saturation. This weak temperature dependence has also recently been measured in samples where good contact to electrodes was achieved by burying the tube ends over a large distance under metallic electrodes fabricated with electron beam lithography \cite{bockrath2, cobden}. It is on the other hand in stark contrast with the T dependence of tubes with tunnel contacts \cite{bockrath,yao} , which exhibit power laws with much larger exponents (of the order of 0.3)  at high temperature, and exponentially increasing resistance at lower temperature because of Coulomb blockade. 
It is not surprising that in addition to the intrinsic conducting properties of tubes (interactions, band structure, disorder), the way they are contacted should determine the temperature dependence of the resistance. For instance  extrapolating the results obtained  on Luttinger liquids between normal reservoirs \cite{safi}, the resistance of a ballistic tube on perfect contacts is expected  to be \textit{insensitive to interactions} and to be given by $R_Q/2=h/2e^2=6.5~k\Omega$ at all temperatures. In contrast, a tube on tunnel contacts will turn insulating at low temperature because of Coulomb blockade. 
Since our measurements are two wire measurements, it is difficult to determine how much of the resistance is due to the contacts, and how much is due to the tube itself. However, the existence in these very same samples of a large proximity effect along with a high supercurrent (see next section) suggests that the transmission at the contacts is close to unity. We therefore attribute the excess resistance compared to the expected value $R_Q/2$ to disorder inside the tubes. 
\begin{table} \centering

\begin{tabular}{|c|c|c|c|c|c|}
\hline
Sample & Contact & length & R(T=290 K) & R(4.2 K)&$l_{e}$ \\ \hline

ST4&TaAu&300 nm&45~$k\Omega$&66 $k\Omega$&50~nm\\ \hline
ST2&TaAu&200 nm&31~ $k\Omega$&33 $k\Omega$&62~nm\\ \hline
ST1&TaAu&300nm&22.1~ $k\Omega$&33 $k\Omega$&90~nm\\ \hline
ST3 & CrIn & 100 nm & 10.7 ~$k\Omega$ & 11.5~$k\Omega$&70~nm \\ \hline
\end{tabular}

\caption{Summary of the characteristics of four SWNT in the normal state.  
The mean free path  $l_e$ is estimated using $R=R_Q~L/(2l_e)$, and assuming a perfect transparency of the contact. The actual mean free path could therefore be larger.\label{tableST} }
\end{table}

\begin{figure}
		\includegraphics[clip=true,width=9 cm]{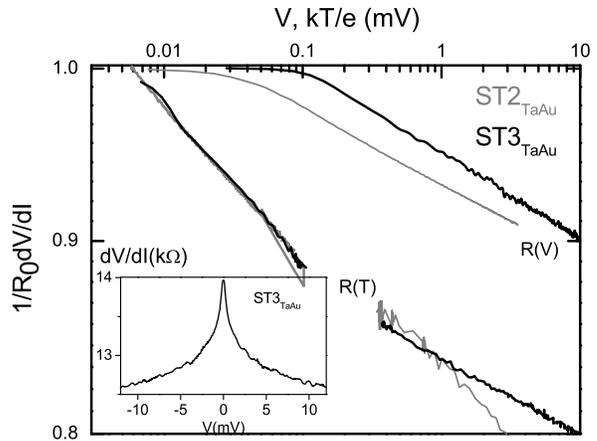}
	\caption{Bias dependence at 50 mK of the differential resistance of individual tubes ST2 and ST3  in comparison with temperature dependence (temperature is expressed in eV units).  \label{RTVswnt}}
	\end{figure} 
	
We then can try to quantify the amount of disorder within our SWNT samples and discuss the possibility of them still being metallic. To do this, we choose to consider the nanotube as a homogeneous diffusive conductor with M conducting channels (in a SWNT M=2), and apply the Drude formula which relates the conductance to the elastic mean free path $l_e$ through $G=M(e^2/h) l_e/L=2G_Q~l_e/L$. The values of $l_e$ deduced in this way are given in Table \ref{tableST}. We find rather short mean free paths compared to the length of the tubes (especially for sample ST4), suggesting the presence of defects in the tubes. The microscopic cause of these short mean free paths could be heptagon-pentagon pairs or defects created by electron irradiation during observations in TEM.

One of the question raised by these small values of $l_e/L$, or equivalently the small values of $G/G_Q$, is why these samples do not localize. Indeed, according to the Thouless criterion \cite{thouless}, a quantum  coherent conductor is expected to localize, i.e. become insulating at low temperature, if its conductance $G$ is  smaller than  $G_Q$. In contrast, we observe no sign of strong localization down to very low temperature where quantum coherence throughout the whole sample is expected to be achieved. 
One possible explanation could lie in the fact that the conductance of a SWNT could be sensitive to interactions even on good contacts as recently suggested\cite{lederer}. These authors, extending the calculations of de Channon \textit{et al.} \cite{dechannon} on chiral  Luttinger liquids to the non chiral case,   predict that the two contact conductance  is renormalized by a factor which  depends on the boundary conditions imposed at the contacts, that can be of the order of the LL  parameter $g$  . In their calculation, there is no nearby gate to screen the interactions, and this is indeed our experimental configuration. These theoretical results could then explain why the high measured resistances do not lead to localization at low temperature. 
However, the same LL theories \cite{grabert2} predict that for a well contacted sample the weak temperature dependence $G-G_0=\alpha T^{-x}$ should saturate at temperatures smaller  than the energy scale  $E_c=hv_\rho/L\sim $ where $v_\rho$ is the plasmon velocity $v_\rho=v_F/g $. In our tubes $E_c$ is of the order of 30~K or more, and we observe no saturation. Therefore we do not think that the low temperature dependence of the resistance observed in these samples contains a clear signature of Luttinger Liquid behavior .

We now examine the voltage dependence of the differential resistance \textit{dV/dI},  below 1 K, and with the contacts in the normal state. Like the temperature dependence of the zero-bias differential resistance, the \textit{dV/dI} has a weak power law dependence as a function of $V$, which can also be fitted by a logarithmic law.  This behavior is at first glance qualitatively similar to what is observed in small metallic bridges presenting a small Coulomb-type anomaly in their differential resistance \cite{weber}. However in the present case the similarity between temperature and bias voltage dependence is only qualitative as shown on Fig.\ref{RTVswnt}: the decrease of differential resistance with $V$ is slower than the corresponding temperature dependence. In addition, the \textit{dV/dI} saturates below  bias values  corresponding to an energy much larger than the temperature. And in contrast to nanotubes connected with tunnel contacts  \cite{bockrath} or small metallic constrictions  \cite{weber}, it is not possible to scale the data using the reduced variable $eV/k_BT$.  This absence of   a voltage  versus temperature scaling  suggests that the temperature dependence of the resistance contains other contributions besides the  expected increase due to electron-electron interactions (for instance weak localization contributions) and calls for further understanding.

\subsection{Ropes of single wall carbon nanotubes}
\begin{figure}
 
\includegraphics[height=8cm]{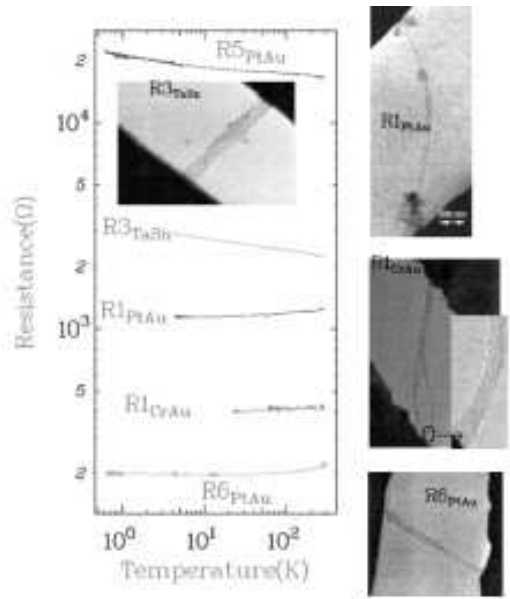}

	%\[	\epsfbox{file.eps}	\]
	\caption{Temperature dependence of the resistance measured on  different ropes  mounted on various types of contacts. For all samples the  measurements were conducted in a magnetic field of 1 T. Note the logarithmic scales.  \label{rtropeshf}}
	\end{figure}

\begin{figure}

\includegraphics[clip=true,width=8cm]{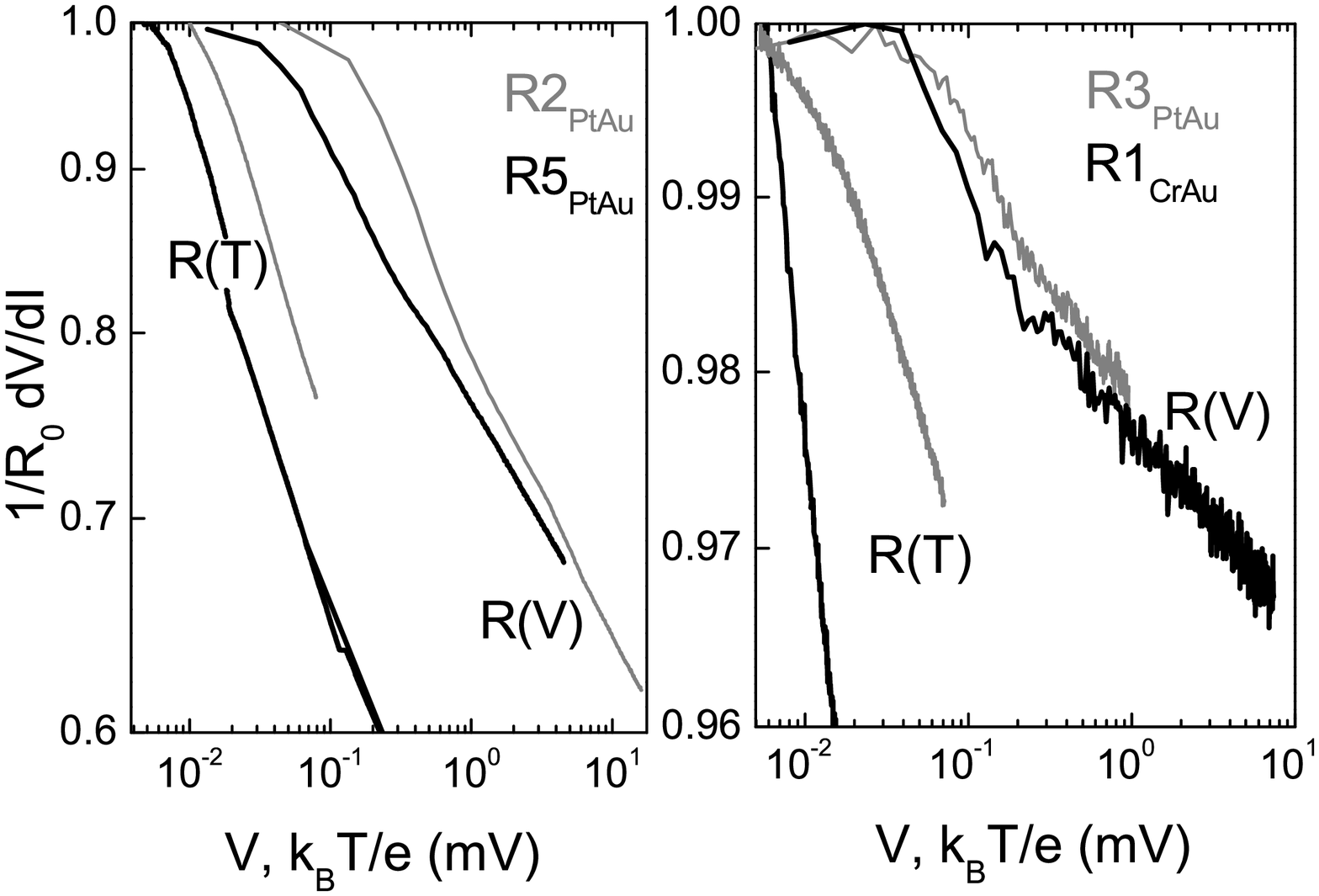}
	%\[	\epsfbox{file.eps}	\]
	\caption{Differential resistance of four ropes in the normal state as a function of voltage bias, and as a function of temperature (expressed in units of voltage). The difference between bias and temperature dependance tends to be weaker  when the resistance of the rope increases. \label{dvirop}}
	\end{figure}

Ropes of SWNT contacted using the same technique  cover a much wider range  of resistances than individual SWNT. The resistances vary between less than $100~\Omega$ and $10^5 ~\Omega$ at 300~K, for ropes containing between forty and several hundred SWNT (see table).  There is no clear correlation between the number of tubes in a rope (deduced from the rope diameter) and the value of its resistance. This may indicate that in some cases only a small fraction of the tubes are connected. In contrast to individual nanotubes, the ropes seem to verify the Thouless criterion: they strongly localize when their resistance is above
$10~k\Omega$ at room temperature and  stay quasi-metallic  otherwise. This behavior is very similar to what is observed in quasi 1D  metallic wires \cite{gershenson}. The temperature dependence of the metallic ropes (see Fig. \ref{rtropeshf}) is also very weak. But for low resistance ropes ($R<10~k\Omega$)  it is not monotonous, in contrast to individual tubes, and as already observed in~\cite{fisher}: the resistance decreases linearly as temperature decreases between room temperature and 30~K indicating the freezing-out of  phonon modes, and then 
increases as $T$ is further decreased, as in individual tubes.  In all samples a bias dependence of the  differential resistance is also observed at low temperature (Fig.\ref{dvirop}). The relative amplitude of these non-linearities are  much smaller than in individual SWNT. This may be related to screening of electron-electron interactions in a rope, which does not exist in a single tube. The magnitude of the resistance variation with $V$ increases with rope resistance, and is weaker than the corresponding temperature dependence. The absence of  scaling laws  in the reduced variable $eV/k_BT$  is  more striking than for individual tubes. In particular, the differential resistance saturates below voltages much larger than temperature.

\begin{table}[t] \centering

\begin{tabular}{|c|c|c|c|c|c|}
\hline
Sample &  length & N (approx.)& R(T=290 K) & R(4,2 K) & 
$l_{e}$\\ \hline
$R1_{PtAu}$  & 1.6 $\mu m$& 300 & 1.1~ $k\Omega$ & 1.2~$k\Omega$ & 40 nm\\ \hline
$R2_{PtAu}$ &$1~\mu m$& 350 & 4.2~$k\Omega$ & 9.2~$k\Omega$&18 nm\\ \hline
$R3_{PtAu}$&$0,3~\mu m$& 240 &475~$\Omega$&500~$\Omega$&9 nm\\ \hline
$R4_{PtAu}$& $1 \mu m$&30&620~$\Omega$&620~$\Omega$&23 nm \\ \hline
$R5_{PtAu}$& 2 $\mu m$&300 & 16~$k\Omega$&21~$k\Omega$& 3 nm\\ \hline
$R6_{PtAu}$&0.3 $\mu m$&200&223~$\Omega$&240~$\Omega$& 30 nm\\ \hline
$R3_{TaSn}$ &0.3 $\mu m$&300&2.2~$k\Omega$&3~$k\Omega$& 3 nm\\ \hline
$R5_{TaSn}$&0.3 $\mu m$&300&700 $\Omega$&1.1~$k\Omega$& 9.2 nm\\ \hline
$R1_{CrAu}$&0.4 $\mu m$&200&400~$\Omega$&$450~\Omega$& 22 nm\\ \hline
$R2_{CrAu}$&0.2 $\mu m$&200&240~$\Omega$&$280~\Omega$& 26 nm\\ \hline
$R1_{ReAu}$&0.5 $\mu m$&200&1750~$\Omega$&$2~k\Omega$& 6 nm\\ \hline
$R2_{ReAu}$&0.5 $\mu m$&200&650~$\Omega$&$700~\Omega$& 18 nm\\ \hline
$R3_{ReAu}$&1.7 $\mu m$&200&65~$\Omega$&$65~\Omega$&572 nm\\ \hline
\end{tabular}
\caption{ Principal characteristics of ropes presented in this 
paper.  The number $N$ of SWNT in each rope is deduced from the diameter of the 
rope.  Mean free paths are estimated assuming that all tubes participate to transport  and that transmission at the contacts is of the order of unity.}
\label{TransnormRswnttab}
\end{table}

\subsection{Importance of intertube coupling in ropes.}

In order to determine whether a rope is closer to an assembly of ballistic tubes in parallel, or to a single diffusive conductor, we need to evaluate the importance of intertube coupling and electron transfer between tubes within a rope.  The characteristic microscopic electron hopping  energy $t_ {\perp}$ between two carbon atoms on different tubes depends on their relative distance $d$ through $t_ {\perp} \propto t_G \exp(-d/a)$. Here \textit{a} is the spatial extension of the $p_{z}$ orbital of carbon, and $t_G$ is the hopping energy between two sheets of graphite \cite{marouf}. Given the differences in C-C distances and orbitals, the ratio between perpendicular and parallel hopping energies in carbon nanotubes has been evaluated to be at most $t_ {\perp } /t_{\parallel} \approx 0.01$.  
In an ordered rope  containing identical individual tubes, authors in \cite{delaney}  have shown that this low intertube coupling can induce a band gap of about 0.1 eV, turning  a rope of metallic tubes into a semiconductor. 

On the other hand in a rope constituted of tubes with different helicities, in the absence of  disorder within the tubes, one finds that the intertube electronic transfer, defined as the matrix element of the transverse coupling between two tubes, integrated over spatial coordinates, is negligeable because of the longitudinal wave function mismatch between neighboring tubes of different helicities. The rope can then be considered as a number of parallel independent ballistic nanotubes. The very  small level of shot noise observed in low resistive ropes  corroborate this statement \cite{roche}. 
However, it has been shown \cite{marouf} that the presence of disorder within the tubes favors intertube scattering by relaxing the  strict orthogonality between the longitudinal components of the wave functions.  Using a very simple model  where disorder is treated perturbatively \cite{ferrier} we find that  for weak disorder  the intertube scattering time is shorter than the elastic scattering time within a single tube.  In tubes longer than the elastic mean free path,  this intertube scattering can provide additional conducting paths to electrons which would otherwise be localized in isolated tubes. Consequently, disordered ropes can be considered as anisotropic diffusive conductors that, in contrast to individual tubes, exhibit a localization length that can  be much larger than the elastic mean free path. This is why we have tentatively  determined for each rope an elastic mean free path  from the value of its conductance using  the  multichannel Landauer expression  for a disordered conductor equivalent to the Drude formula:
$G=2M(e^2/h) l_e/L$
with the number of channels $M=2N_t$ where $N_t$ is the number of tubes inside the sample. Since the number of metallic tubes connected is statistically of the order or less than $N_t/3$ these values of $l_e$ are   certainly underestimated. In addition, it may be possible that the metallic tubes in a rope are not all well connected to both electrodes. Then transport through the rope will proceed via inter tube transfer. That is why ropes with many disordered metallic tubes might have very high resistances.

\section{Proximity induced superconductivity}

A non-superconducting or normal metal (N) in  good contact with a macroscopic superconductor (S) is in the so-called 'proximity effect regime': superconducting correlations enter the normal metal over a characteristic length $L_N$ which is  the smallest of either the
phase coherence length in the normal metal $L_{\phi}$ or the thermal length
$L_T$, (in a clean metal $L_T=\hbar v_F/k_B T$ and in a dirty metal
$L_T=\sqrt{\hbar D/k_BT} $ where $D$ is the electron diffusion coefficient). Both 
lengths, of the order of a few $\mu m$, can be much longer than the
superconducting coherence length $\xi=hv_F/\Delta$ or $\sqrt{\hbar D/\Delta}$ where $\Delta$ is the energy gap of the superconducting contacts.

If the normal metal's length is less than $L_N$ and if the resistance of the NS interface is sufficiently small, a gap in the density of states is induced in the normal metal, that  is as large as the gap of the superconductor in the vicinity of the interface and decreases on a typical lengthscale of $L_N$. Consequently, a normal metal shorter than $L_N$ between two superconducting electrodes (an SNS junction) can  have a critical temperature (equal to that of the superconductor alone) and exhibits a Josephson effect, i.e.  a supercurrent at zero bias. This manifestation of the
proximity effect has been extensively studied in multilayered planar SNS
junctions \cite{deutscher}  and more recently in lithographically fabricated
micron scale metallic wires made of normal noble metals between two
macroscopic superconducting electrodes \cite{courtois,dubos}. The  maximum
low temperature value of the supercurrent (critical current) in such SNS
junctions  of normal state resistance $R_N$ is  : $ \pi\Delta/e R_N$ in the short junction limit $L\ll \xi$ or  $\Delta \ll E_c$\cite{lik},  and $ \alpha E_c/e R_N$ in the limit of long junctions $L\gg \xi$ or  $E_c \ll \Delta $ \cite{dubos}. Here $\alpha$ is a numerical factor of the order of 10, and $E_c$ is the Thouless energy $E_c=\hbar D/L^2$.

Probing the proximity effect in a normal wire connected to superconducting electrodes and in particular the existence of a Josephson current constitutes a  powerful tool for the investigation of phase coherent transport through this normal wire. In the next paragraph we present experimental results on carbon nanotubes mounted on superconducting electrodes showing evidence of large supercurrents which indicate that coherent transport takes place on micron length-scales.

\subsection{Proximity effect in individual tubes}

      We  have observed  proximity induced
superconductivity in the three different individual SWNT  ST1, ST2 and ST4 described in the previous section.  These samples are  mounted on  Ta/Au electrodes  see table \ref{tablecontacts} which are a bilayer
(5 nm Ta, 100 nm Au) with a transition temperature of the order of  0.4 K. This value is strongly reduced compared to the transition temperature of bulk tantalum (4 K) due to the large  thickness of gold relative to tantalum.

\subsubsection{Zero bias resistance}

\begin{figure}
\includegraphics[clip=true,width=9cm]{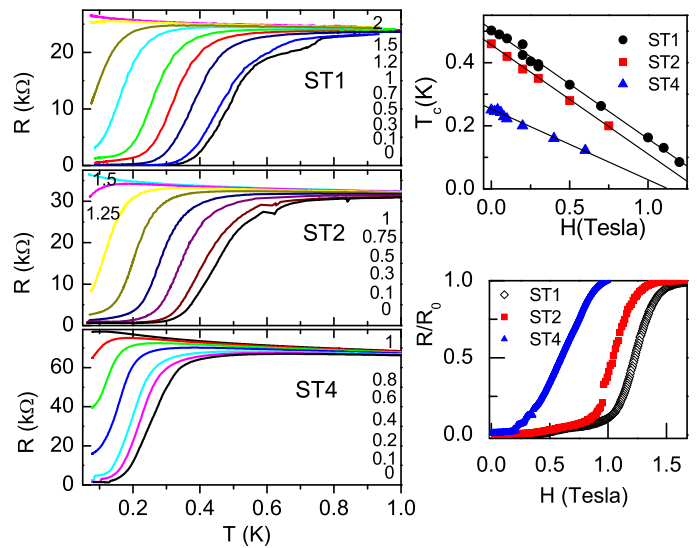}

	%\[	\epsfbox{file.eps}	\]
	\caption{Left panel:  Temperature dependence of the resistance of 3 different single tubes $ST1,2,4.$ mounted on TaAu, for different values of the magnetic field perpendicular to the tube axis in the plane of the contacts. (The labels on the curves correspond to the value of magnetic field in Tesla). For some samples the resistance of the contacts exhibited several steps in temperature, probably indicating inhomogeneities in the thickness of the gold layer. This may explain the small resistance drop observed on sample ST1 around 1 K at low field. Top right panel: field dependence of the transition temperature, defined as the inflection point of  $R(T)$. Bottom right  panel:  Magnetoresistance  of the  single tubes measured at 50 mK (the magnetic field is perpendicular to the tube axis in the plane of the contacts). This positive magnetoresistance only saturates at very high field (2T). \label{rhst}}
	\end{figure}

For these three samples  the zero bias resistance exhibits a broad transition around the superconducting transition temperature of the contacts and becomes zero at lower temperature except for the highest resistance sample ST4 ($R_N=70~k\Omega$) which has a residual resistance of 800~$\Omega$.  The transition is
shifted to lower temperature when a magnetic field is applied in the plane of
the contacts and perpendicular to the tube axis. Above 2 T  the resistance is field independent and slightly increases when the temperature is lowered below $0.2~K$ as already mentioned in the previous section . The critical field is of the order of 1 T and can  be also extracted as the inflection point of the magnetoresistance  depicted in Fig. \ref{rhst}. This value is surprisingly high since it is ten times  larger than the measured
critical field of the contact (0.1 T).   This high value of critical field  could be due to local modifications of the bilayer  Ta /Au  film in the contact  region due to the laser pulse, in particular the melted upper gold film is probably much thinner than the original one.
It is however important to note that these critical fields are of the same order of magnitude  in all the samples measured. As  Fig. \ref{rhst} shows, the transition line $T_c(H)$ is a linear function of filed for all samples.

\subsubsection{Bias dependance and critical current}

\begin{figure}
	\includegraphics[clip=true,width=8cm]{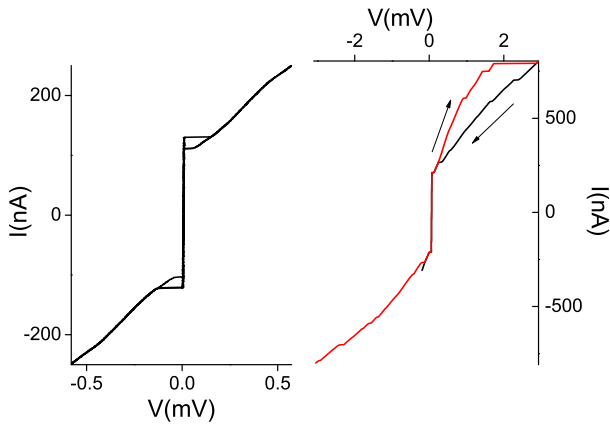}
		%\[	\epsfbox{file.eps}	\]
		\caption{$I$ Vs $V$ curve of the single tube $ST1$, from which the critical current is deduced. Right panel, data on a wider current scale showing the existence of voltage steps for currents above $I_c$. The arrows indicate the current sweep direction. \label{ivst}}
		\end{figure}

The most striking signature of induced superconductivity  is the existence of Josephson 
supercurrents through the samples \cite{Kasumov}. The existence of a supercurrent shows up in the voltage vs. current curves and differential resistance, see Fig. \ref{ivst}. For sample $ST_1$,  the
transition between the superconducting state (zero voltage drop through the
sample) and the dissipative (resistive) state   is quite abrupt and displays hysteresis at low
temperature. It is characterized by a critical current $I_c=0.14~\mu A$ near zero temperature.  Similar behavior with  smaller values of critical current is also observed in the other samples (see table  \ref {proxtab1}). The product $R_N I_c$ at $T\approx0$, , varies between 1.6 and 3.5 mV. If we  deduce $\Delta$ from the $T_c$ of the superconducting contacts, we find that $R_N I_c$   is more than \emph{ten times larger}
than the  maximum expected  value ($ \pi\Delta/e  =0.2~mV$) for the short junction limit \cite{lik}, and all  samples are in this short junction limit $E_c >\Delta$. It is  then  also interesting to note that the product $ e R_N I_c$ is closer to the gap of pure tantalum $\Delta_{Ta}= 0.7~ meV $. But it is difficult to understand why  the induced gap could be the gap of pure tantalum whereas the resistance  drop of the tube follows the transition of the Au/Ta bilayer.

\begin{table}[htb] %\centering

\begin{tabular}{|c|c|c|c|c|c|c|c|}
\hline
 &length & $T_{C}$ & $R_{N}$ & $I_{C}$ & $\Delta_{contact} $& $R_{N}I_{C}$ 
 &$E_{C} $ \\ \hline

ST1 & 0.3~$\mu m$ & 0.5 K & 25 $k\Omega$ &0.14~$\mu A$ & 0.07 & 3.5 
mV& 0.4 \\ \hline
ST2 & 0.3~$\mu m$ & 0.45 K & 33 $k\Omega$ & 0.075~$\mu A$ & 0.07 & 2 
mV& 0.3 \\ \hline
ST4 & 0.3~$\mu m$ & 0.25 K & 65 $k\Omega$ & 0.025~$\mu A$ & 0.07 & 
1.6 mV& 0.16 \\ \hline

\end{tabular}
\caption{Principal features of  the superconducting junctions obtained with individual tubes.   
$T_{C}$ is the transition temperature of the contact-SWNT-contact junction,  $R_{N}$ is the normal state resistance  and 
$I_{C}$ is the critical current of the junction.   $\Delta_{contact}$ 
is the gap of the contacts estimated from their transition temperature. The Thouless energy 
$E_{C}$ is computed using the mean free paths $l_{e}$ estimated in the 
previous section. These energies are expressed in meV.}
\label{proxtab1}
\end{table}

Somewhat unexpectedly and uncharacteristic of metal SNS junctions, the normal state resistance is not recovered above the
critical current, but the V(I) curve shows further hysteretic jumps  at higher currents. 

\begin{figure}
	\includegraphics[clip=true,width=9 cm]{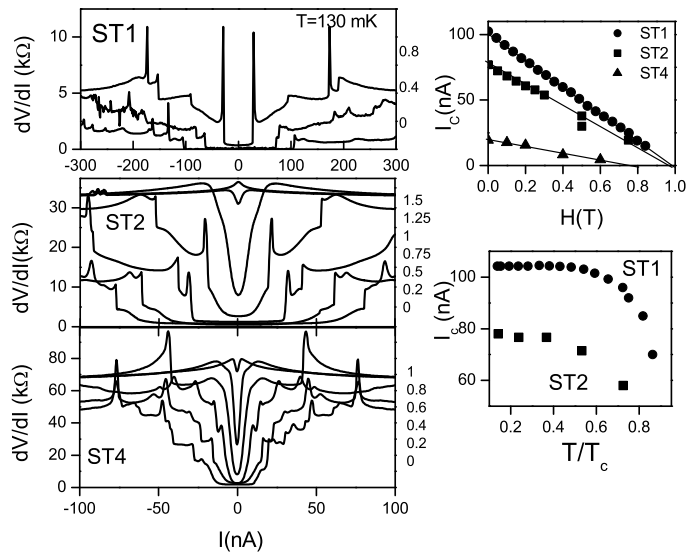}
				%\[	\epsfbox{file.eps}	\]
		\caption{Differential resistance  of the 3 samples ST1, 2 and 4 for different
magnetic fields.  Right panels: Field and temperature dependence of the critical current, defined as the current at which the first resistance jump occurs. The critical current disappears around $H_c=1$ ~T, which is also the value deduced from the low temperature $R(H)$ curves in Fig. \ref {rhst}.\label{didvst}}
		\end{figure}

These features also appear in the differential resistance  $dV/dI$ (see Fig. \ref {didvst}). The superconducting state corresponds to the  zero differential resistance at  low dc current. At the critical current  the differential resistance displays a sharp peak  followed by smaller ones at higher current.
Each peak corresponds to a hysteretic feature in the  dc $V-I$ curve . The peaks are linearly shifted to
 lower current when increasing magnetic field and all disappear above 2T. The temperature and field dependencies of the critical current extracted from this data are plotted in Fig. \ref{didvst}.  The temperature dependance  is very  weak below $0.8~T_c$ and decreases rapidly above. It does not follow the behavior expected for SNS junctions (neither in the limit of short junctions nor long junctions where an exponential decay of the critical current is expected \cite{dubos} ). 
The field dependance of the critical current is precisely linear for all samples, with disappearance of critical current above 1 T.

It is very difficult to understand these results in the framework of conventional proximity induced superconductivity. In particular we have already mentioned that the IV curves exhibit non linearities, and signs of superconductivity at very large bias,  i.e. much larger than the gap of the Ta/Au contacts (and even the gap of pure tantalum!). Such non-linearities recall manifestations of phase slips in 1D superconductors\cite{meyer}: Above $I_c$, small  normal regions of size comparable to the inelastic length $L_N$ are nucleated around
defects in the sample (phase slip centers). They have not to our knowledge been observed before in SNS junctions. These observations by themselves suggest the possibility of intrinsic superconducting correlations in SWNT, which will be discussed further on.

It is therefore interesting to compare these results with the superconductivity of other types of metallic nanowires. Ref.\cite{Bezryadin} reports superconducting nanowires fabricated by depositing   thin films of amorphous MoGe  on suspended carbon nanotubes.  Disappearance of superconductivity in the most resistive wires  is attributed to the formation of quantum phase slips at very low temperature. The first  data  obtained on these samples indicated disappearance of superconductivity for normal state resistance larger than $h/4e^2$ .  In a later paper \cite{tinkham2}, the authors show that what determines the suppression of superconductivity is not the normal state resistance but rather the resistance per unit length. They find that the transition is suppressed for resistances per unit length larger than roughly 0.1 k$\Omega$/nm. We have observed proximity induced superconductivity in SWNTs with a normal state resistance up to 70~k$\Omega$,  with a resistance per unit length of the order of 0.25 k$\Omega$/nm. Note however that the MoGe nanowires  contain typically 10000 channels, whereas our nanotubes only contain 2 conducting channels. This comparison tells us that proximity induced superconductivity is amazingly robust in our samples.

Finally signs of proximity induced superconductivity have also been observed by other groups, on individual SWNT \cite{morpurgo} and on multiwall nanotubes \cite{buitelar}, but without supercurrents. In appendix A of this paper we show our results on proximity-induced superconductivity in multiwall nanotubes.

\subsection{Proximity induced superconductivity in ropes}
\begin{figure}
\includegraphics[clip=true,width=8.5cm]{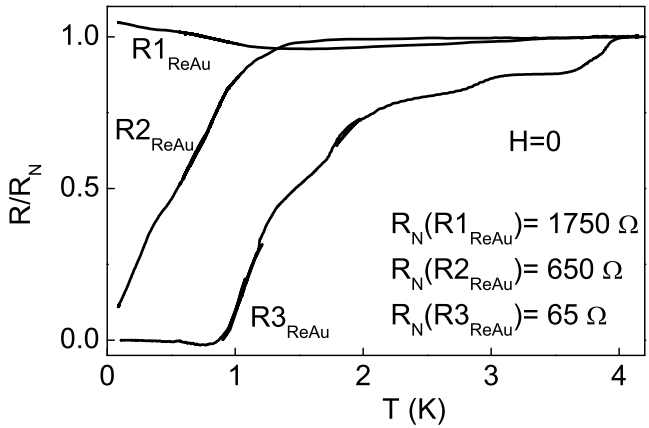}
	%\[	\epsfbox{file.eps}	\]
	\caption{ Temperature dependence of the resistance of ropes
	$R1_{ReAu}$, $R2_{ReAu}$, $R3_{ReAu}$ mounted on Re/Au contacts, in zero magnetic field. The $T_c$ of Au/Re is 1 K. Sample 
$R3_{ReAu}$ becomes superconducting below 1K, but the resistance steps above the transition indicates  that the superconductivity of the contacts may not be homogeneous. \label{rtropes}}
\end{figure}

\begin{figure} 
 \includegraphics[clip=true,width=8.5cm]{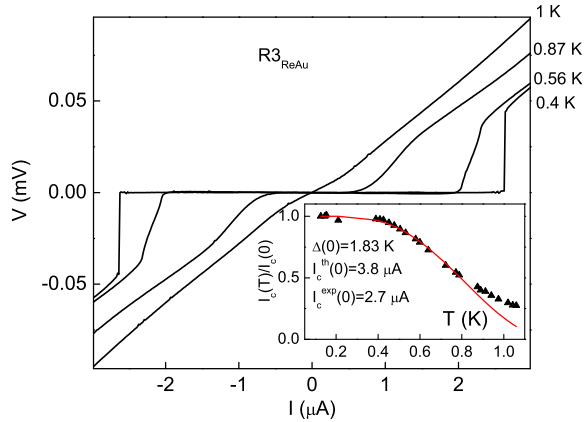}
	     %\[ \epsfbox{file.eps}	\]
	 \caption{Temperature dependence of the critical current of rope $R3_{ReAu}$ . Main panel: I-V curves at different temperatures. Inset: Temperature dependence of the critical current (triangles), and fit to the Ambegaokar Baratoff formula $i_{AB}(T)=i_0\frac{\Delta(T)}{\Delta(0)}\tanh(\Delta(T)/2k_B T)$ (continuous line). \label{ictrope}}
	 \end{figure}

\begin{figure}
\includegraphics[clip=true,width=8cm]{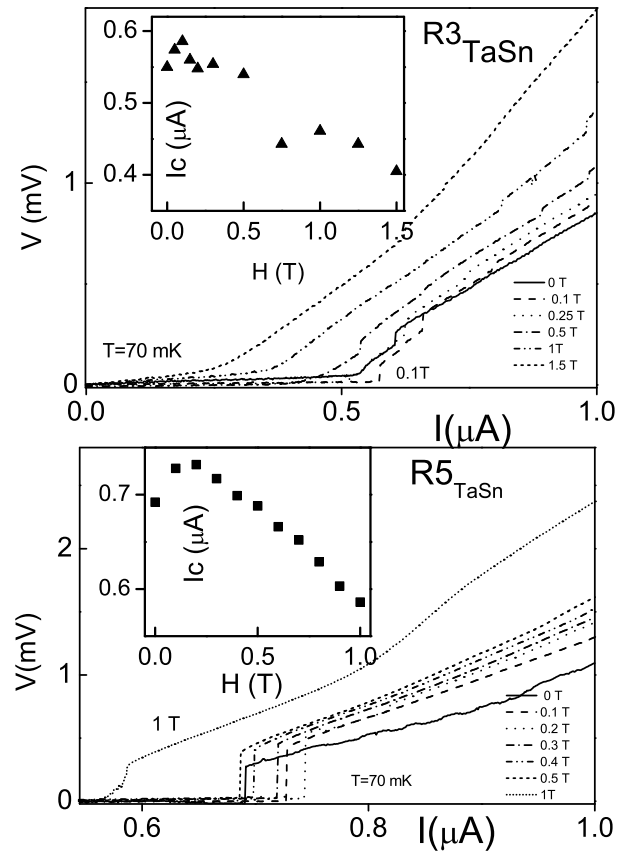}
				%\[	\epsfbox{file.eps}	\]
		\caption{$V$ versus $I$ curves of ropes $R3_{TaSn}$ and  $R5_{TaSn}$ in fields ranging from 0 to 1.5~T. Inset: field dependence of the critical current deduced form these IV curves. The field is parallele to the contacts, and perpendicular to the tubes. Note the non monotonous behavior. \label{ivropes}}
		\end{figure}

We also found proximity induced superconductivity in ropes of SWNT mounted on Re/Au or Ta/Sn  which characteristics are shown in table \ref{proxmr} and Fig. \ref{rtropes} to \ref{ivropes}. 

\begin{figure}
\includegraphics[clip=true,width=9cm]{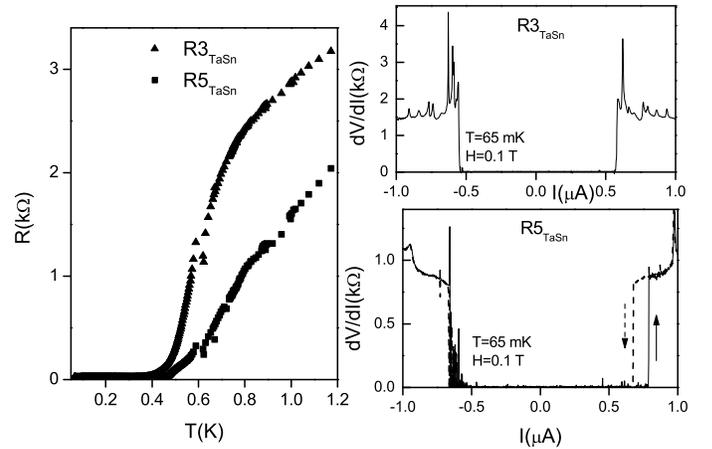}
\caption{Left panel: Temperature dependence of zero-bias resistance of ropes $R3_{TaSn}$ and  $R5_{TaSn}$. Right panels: differential resistance curves taken at 0.1 T, with an example of the hysteresis with current sweep direction.}
\end{figure}

\begin{table}[htb] %\centering

\begin{tabular}{|c|c|c|c|c|c|c|c|}
\hline
 &length& $T_{C}$ & $R_{N}$ & $I_{C}$ & $\Delta$& $R_{N}I_{C}$ 
 &$E_{C}$ \\ \hline
$R1_{ReAu}$ & 0.5~$\mu m$ &$\leq50~mK$ & 1750 $\Omega$ & 0 & 0.12   & 0& 0.01 
  \\ \hline
$R2_{ReAu}$ & 0.5~$\mu m$ & $\leq 50~mK$ & 650 $\Omega$ & 0 & 0.12   & 0 & 0.03 
  \\ \hline
$R3_{ReAu}$ & 1.7~$\mu m$ & 1 K & 65 $\Omega$ &2.7$ \mu A$ & 0.12   & 0.175 
mV& 0.13   \\ \hline
$R3_{TaSn}$ &  0.4~$\mu m$ & 0.7 K & 2200 $\Omega$ &0.54 $ \mu A$ & 0.6  & 1.5 
mV& 0,01   \\ \hline
$R5_{TaSn}$ & 0.4~$\mu m$ & 0.6 K & 700 $\Omega$ &0.77 $ \mu A$ & 0.6   & 0.63 
mV& 0.03   \\ \hline
\end{tabular}
\caption{Main features of the ropes suspended between superconducting contacts.  
$T_{C}$ is the transition temperature of the junction, while $\Delta$ 
is the gap of the contacts.  $R_{N}$ is the resistance at 4.2 K and 
$I_{C}$ is the critical current of the junction.  Thouless energy 
$E_{C}$ is computed using the mean free paths $l_{e}$ computed in the 
preceding section. Energies are in meV.}
\label{proxmr}
\end{table}
	
It is interesting to note that in contrast to what was found in SWNTs, we only observe proximity-induced superconductivity in ropes with a normal state resistance less than $10~k\Omega$. More resistive ropes, as previously mentioned, are insulating at low temperature. The data of Fig. \ref{rtropes} for ropes mounted on Re/Au  show that although a resistance drop is observed in the vicinity of the transition of the contacts, proximity induced superconductivity is not always complete and is only slightly visible on the most resistive sample. This may result from the combined effects of the quality of the contacts and disorder in the rope for which the condition $ L_N >L$ may not be fulfilled in our available temperature range. 
The critical current deduced at 50~mK from the I-V curves are given in table \ref{proxmr} and compared to the theoretical predictions. The rope on Re/Au contacts is in the short junction regime  ($\Delta < E_c$) and the ropes on TaSn are in the long junction limit.   In all cases  the measured supercurrents are  larger than allowed by the theory of SNS junctions.  It is noteworthy that the two ropes mounted on tin have similar critical currents ($I_c\approx0.7~\mu A$) in spite of the fact that their  normal state resistances differ by  more than a factor three.   As for the individual tubes, jumps in the IV curves are also observed  at bias larger than the gap of tin. The temperature dependence of the critical current  in $R3_{ReAu}$ (see Fig. \ref{ivropes}) is more pronounced than for individual tubes, and can be approximately fitted by an Ambegaokar-Baratoff law typical of tunnel junctions \cite{Amb}. 
In the case of ropes mounted on tin contacts this temperature dependence is difficult to extract  because of the rounding of the IV curves with  increasing temperature. 
In contrast to individual tubes, the field dependance is not monotonous. The critical current first increases up to 0.1 T where it goes through a maximum and decreases linearly at higher field up to 1 T.  Such non-monotonous behavior  has been predicted  in a number of models  describing non homogenous superconductors and could be related  to interference between different electronic trajectories along the various tubes of the rope and  the existence of negative Josephson couplings between these tubes. This recalls the non-monotonous magneto-resistance of amorphous superconducting wires\cite{dynes}  and in MWNT (see appendix), and may have a common origin. 

\subsubsection*{Effect of magnetic field orientation}

\begin{figure}
\includegraphics[clip=true,width=8cm]{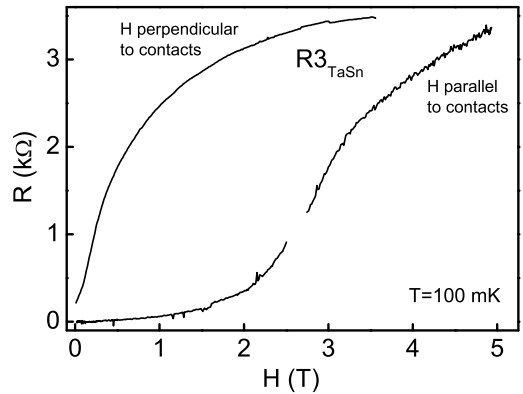}
	%\[	\epsfbox{file.eps}	\]
	\caption{  Magnetoresistance  of  the  rope $R3_{TaSn}$  measured at 50 mK for the field parallel and perpendicular to the contacts, and always perpendicular to the tube. \label{rhrope}}.
	\end{figure}

\begin{figure}
\includegraphics[clip=true,width=8cm]{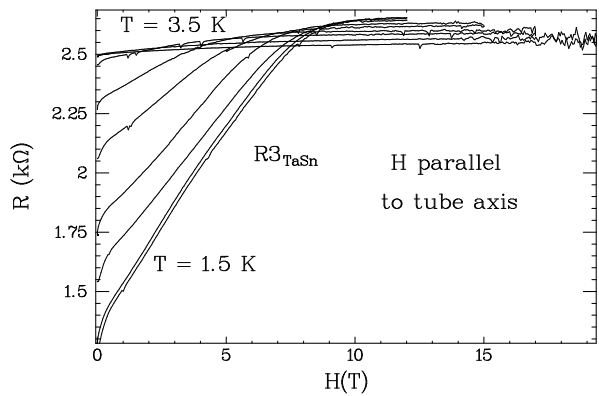}
	%\[	\epsfbox{file.eps}	\]
	\caption{Magnetoresistance  of  the  rope $R3_{TaSn}$  measured at 1 K with the field along the tube  in the high field facility of Grenoble. Note that the magnetoresistance increases in magnetic field up to H=10 T. \label{rhrope2}}
	\end{figure}

For the ropes mounted on tin we investigated the magneto-transport for different field orientations. We found that with the field parallel to the contacts (and perpendicular to the tubes) the critical fields were much larger than the critical field of the contacts. This may be explained by a local  thinning of the metallic layer contacting the tube in the contact region after soldering. Figure \ref{rhrope} shows the magnetoresistance for these 2 different field orientations.  For these samples the magnetoresistance  was  also measured  between 0 and 20 T using the high field facility in Grenoble at 1 K with the magnetic field  along the rope (see  fig. \ref{rhrope2}).   The magnetoresistance increases linearly with field and saturates only above 10 T. This field is of the order of the Clogston Chandrasekhar \cite{Clogston} criterion for destruction of superconductivity   by  pair breaking due to spin polarization: $\mu_{0}H_{p}=\Delta _{Sn} /\mu_{B}\approx7$~T, where $\Delta _{Sn}$ is the gap of tin. 
Unfortunately, the high field experiments were not conducted in the  same temperature range as the low field ones, and in the second measurement (low field), because of thermal cycling, the sample exhibited less proximity effect. Nevertheless, it seems as though the superconductivity depends on the orientation of the magnetic field with respect to the tube and not only on its orientation  with respect to the contacts, as expected for proximity induced superconductivity.

\subsection{Discussion}
Observation of a strong proximity effect indicates that phase coherent transport takes place in carbon nanotubes on the micron scale. However in the case of single wall tubes  the  surprisingly high values of critical currents cannot be described by the standard theory of SNS junctions.
The problem of an SNS junction constituted of  a Luttinger liquid between two
superconducting reservoirs has been considered theoretically 
 for weakly  transmitting  contacts \cite{fazio}, and for perfectly transmitting contacts\cite{maslov,saleur}. In both limits the
authors predict that it is indeed possible to induce superconductivity by a
proximity effect. In the case of repulsive interactions   and perfectly
transmitting interfaces the zero temperature value of the critical current is 
not changed by the interactions. More recently it has been shown that the presence of attractive interactions in a Luttinger liquid can result in a significant increase of the critical current \cite{affleck,gonzalez} and unusual temperature dependencies very similar to what we observe for individual tubes. A more accurate fit  of the temperature dependance of the critical current  in the rope  including the existence of the inflection point   can also be obtained using the theoretical predictions of Gonzalez  \cite{gonzalez} .

Our data could also be explained by the existence of superconducting
fluctuations intrinsic to SWNT.  For an infinite nanotube,  because of its 1D
character, these fluctuations are not expected to give rise to a
superconducting state at finite temperature. However, the superconducting state
could be  stabilized by the macroscopic superconductivity of the contacts. In such a situation, it is conceivable to expect the critical current  to be enhanced compared to its value in a conventional SNS  junction and to be given by the critical current of a superconducting 2-channel wire: $I_c= (4e^2/h) \Delta_t $  \cite{klap} determined by the value of the  superconducting pairing amplitude $\Delta_t $ inside the wire and independent of the normal  state resistance of the nanotube (in the limit where the mean free path is larger than the superconducting coherence length). The existence
of  superconducting fluctuations intrinsic to nanotubes may  also  help to explain the
positive magneto-resistance  observed in all our samples where the
normal state resistance is recovered at fields much higher than the critical
field of the contacts. In the next section we show that these hypotheses are corroborated by the observation of intrinsic superconductivity in ropes of SWNT. 
\section{Intrinsic superconductivity in ropes of SWNT on normal contacts}

\begin{table}[bp] %\centering
\begin{tabular}{|c|c|c|c|c|c|c|c|}
\hline
 &L & N & $R_{290K}$ & $R_{4.2K}$&T*&$I_c$&$I_{c}$$^*$ \\ \hline
$R1_{PtAu}$ & $2~\mu m$ & 350 & 10.5 $k\Omega$ & 1.2 $k\Omega$ &140 mK & 
0.1 $\mu A$ &0.36 $\mu A$ \\ \hline
$R2_{PtAu}$ &$1~\mu m$& 350 & 4.2$k\Omega$ & 9.2$k\Omega$&550 mK&0.075 $\mu 
A$&3 $\mu A$\\ \hline
$R3_{PtAu}$&$0.3 \mu m$& 350 &400 $\Omega$&450 $\Omega$&*&*&*\\ \hline
$R4_{PtAu}$&$1 \mu m$&45&620 $\Omega$&620 $\Omega$&120 mK&*&$0.1~\mu A$ \\ \hline
$R5_{PtAu}$&2 $\mu m$&300&16 $k\Omega$&21 $k\Omega$&130 mK&20 $n A$&0.12 $\mu A$\\ \hline
$R6_{PtAu}$&0.3 $\mu m$&200&240 $\Omega$&240 $\Omega$&*&*&*\\ \hline

\end{tabular}
\caption{Summary of the characteristics of six ropes mounted on Pt/Au contacts.  T* is the temperature below which the resistance starts to drop, $I_c$ is the current at which the first resistance increase occurs, and $I_{c}$$^*$ is the current at which the last resistance jump occurs.}
\label{supratab}
\end{table}

\begin{figure}[h]

\begin{center}
\includegraphics[clip=true,width=9.5cm]{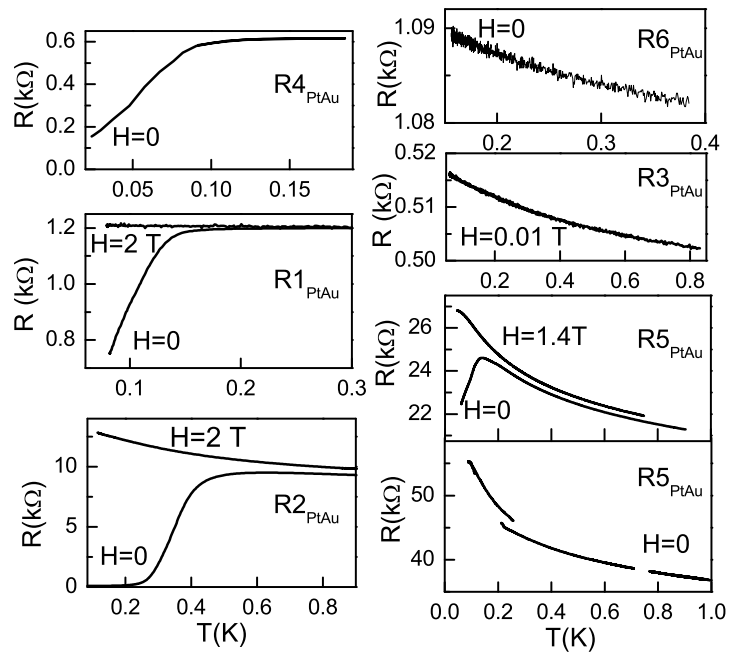}
%\leavevmode
%\epsfxsize=8cm
%\epsfbox{fig1.eps}
\end{center}
%\vspace{0 cm}
\caption{
\label{figurert}
Resistance as a function of temperature for the  six samples described in table VI, both in zero fields and large fields.}

\end{figure}

In the following we discuss the low temperature transport (below 1K) of suspended \textit{ropes} of SWNT connected to normal electrodes. The electrodes are trilayers of sputtered $Al_{2}O_{3}/Pt/Au $ of respective thickness 5, 3 and 200 nm. They do not show any sign of superconductivity down to 50 mK. 
 
\begin{figure}[h]
\begin{center}
\includegraphics[clip=true,width=8cm]{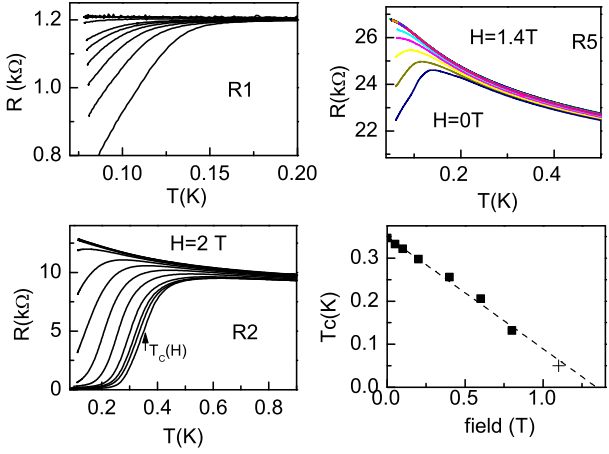}
\end{center}
%\vspace{0 cm}
\caption{
\label{figurerth}
Resistance as a function of temperature for  samples  $R1,2,5_{PtAu}$ showing   a transition. The  resistance of R1 is measured in magnetic fields of $\mu_{0}H$= 0, 0.02, 0.04, 0.06, 0.08, 0.1, 0.2, 0.4, 0.6, 0.8 and 1 T from bottom to top.  The resistance of R2 is taken at $\mu_{0}H$=0, 0.05, 0.1, 0.2, 0.4, 0.6, 0.8, 1, 1.25, 1.5, 1.75, 2, 2.5 T from bottom to top. That of R5 at $\mu_{0}H$=0, 0.1, 0.2, 0.3, 0.5, 1.4 T from bottom to top. Bottom right: $T_c(H)$ for R2.}
\end{figure}

\noindent

As shown in figure \ref{figurert}, different behaviors are observed for the temperature dependence of the zero bias resistance. The resistance of some samples ($R3_{PtAu}$ and $R6_{PtAu}$) increases weakly and monotonously as T is reduced, whereas the resistance of others ($R1,2,4,5_{PtAu}$) drops over a relatively broad temperature range, starting below  a temperature $T^*$  between 0.4 and 0.1 K ($T_{1}^*= 140$ mK, $T_{2}^*= 550$ mK,  $T_{4}^*= 100$ mK).  The resistance of $R1_{PtAu}$ is reduced by 30\% at 70 mK and that of $R4_{PtAu}$  by 75\% at 20 mK.  In both cases  no inflection point in  the temperature dependence is observed.   On the other hand the resistance of $R2_{PtAu}$ decreases by more than two orders of magnitude, and reaches a constant value below 100 mK  $R_{r}=74$~$\Omega$. This drop of resistance disappears when increasing the magnetic field. For all the samples we can define a critical field  above which the normal state resistance is recovered.  As shown on Fig. \ref{figurerth} this critical field decreases linearly with  temperature, very similar to what is seen in SWNT and ropes connected to superconducting contacts. We define a zero temperature critical field $H_c$ as the  extrapolation of $H_c(T)$ to zero temperature (see Fig. \ref{figurerth}). 

\begin{figure}[hbt]
\begin{center}
\includegraphics[clip=true,width=9cm]{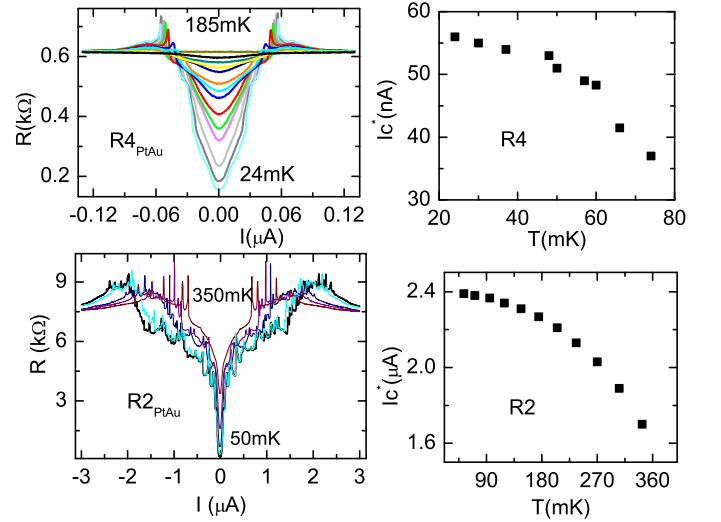}
\end{center}
\caption{
\label{figuredivt}
Differential resistance of $R2 _{PtAu}$ and $R4 _{PtAu}$, at different temperatures. Right panel: Temperature dependence of $I_{c}$$^*$, the current at which the last resistance jumps occur in the $dV/dI$ curves.}

\end{figure}

\begin{figure}[hbt]
\begin{center}
%\vspace{1 cm}
%\leavevmode
%\epsfxsize=7.0cm
%\epsfbox{fig2.eps}
\includegraphics[clip=true,width=9cm]{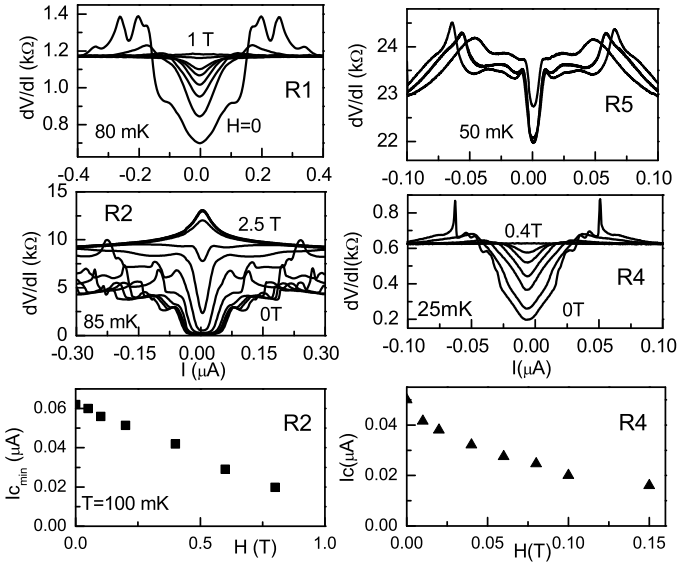}
\end{center}
\caption{
Differential resistance as a function of current for samples $R1,2,4,5_{PtAu}$ in different applied fields. Sample R1: Fields are 0, 0.02, 0.04, 0.06, 0.08, 0.1, 0.2 and 1 T. Sample R2: Fields are 0, 0.2, 0.4, 0.6, 0.8, 1, 1.25, 1.5, 1.75, 2, and 2.5 T. Sample R5: Fields are 0.02, 0.04, 0.06, 0.08 T. Sample R4: Fields are 0, 0.02, 0.06, 0.1, 0.15, 0.2 and 0.4 T. Bottom: Field dependence of $I_{c}$ for  samples $R2 _{PtAu}$ and $R4 _{PtAu}$. Note the linear behavior.\label{figuredivh}}
\end{figure}

Above the critical field, the  resistance increases with decreasing temperature, similarly to  ropes 3 and 6, and becomes independent of magnetic field. Fig. \ref{figuredivt} and Fig. \ref{figuredivh} show that in the temperature and field range where the zero-bias resistance drops, the differential resistance is  strongly current-dependent, with lower resistance at low current. These data suggest that the ropes 1, 2 and 4 are superconducting. Although the experimental curves for $R2_{PtAu}$  look similar to those of SWNTs connected to superconducting contacts \cite{Kasumov},  there are major differences. In particular the $V(I),~dV/dI(I)$ do not show any supercurrent because of the existence of a finite residual resistance due to the contacts being normal.

Before analyzing the data further we wish to emphasize that 
this is the first observation of superconductivity in wires having less than one hundred conduction channels.
Earlier experiments in nanowires \cite{Giordano,dynes,tinkham2} dealt with at least a few thousand  channels. We therefore expect a strong 1D behavior for the transition. In particular, the broadness of the resistance drop with temperature is due to large fluctuations of the superconducting order parameter in reduced dimension starting at the 3D transition temperature $T^*$.
In the following we will try to explain the variety of behaviors observed  taking into account several  essential features: the large normal contacts,  together with the finite length of the samples compared to relevant mesoscopic and superconducting scales,  the number of tubes within a rope, the amount of disorder  and intertube coupling.  We first assume  that all ropes are diffusive conductors but we will see  that this hypothesis is probably not   valid in the more ordered ropes.

\subsection{Normal contacts and residual resistance}

We first recall that the resistance of any superconducting wire measured through normal contacts (an NSN junction) cannot be zero : a metallic SWNT, with 2 conducting channels, has a contact resistance of half the resistance quantum, $R_{Q}/2$ (where $R_{Q}=h/(2e^2)$=12.9 k$\Omega$), even if it is superconducting. A rope of  $N_{m}$ parallel metallic SWNT will have a minimum resistance of $R_{Q}/(2N_{m})$. Therefore we use the residual resistance $R_{r}$ to deduce a  lower bound for the number of metallic tubes in the rope  $N_{m}=R_{Q}/2R_{r}$.  From the  residual resistances  of $74~ \Omega$ in sample $R2_{PtAu}$, and  less than $170~\Omega$  in $R4_{PtAu}$ we deduce that there are  at least $\approx 90$ metallic tubes in  $R2_{PtAu}$  and  $\approx40$ in $R4_{PtAu}$.
In both cases that means that a large fraction of tubes participate to transport and justifies a posteriori the hypothesis that ropes are diffusive conductors. In the other samples we cannot estimate the number of conducting tubes because we do not reach the regime where the resistance saturates to its lowest value.

\subsection{Estimate of the superconducting coherence length.}

Assuming that the BCS relation holds we get for the superconducting gap $\Delta =1.76$ $k_{B}T^*$ : $\Delta \approx 85$~$\mu eV$ for $R2_{PtAu}$. We can then deduce the superconducting coherence length along the rope in the diffusive limit :
\begin{equation}
\xi_=\sqrt{\hbar v_{F}l_{e}/\Delta}
\label{XI}
\end{equation}
This expression yields  $\xi_{2}\approx 0.3$~$\mu$m where $v_{F}$ is the longitudinal Fermi velocity $8\times10^5$ m/s.(Consistent with 1D superconductivity, $\xi_{2}$ is ten times larger than the diameter of the rope). 
We now estimate the superconducting coherence length of the other samples, to explain the extent or absence of observed transition. Indeed, investigation of the proximity effect at high-transparency NS interfaces has shown that superconductivity resists the presence of normal contacts only if the length of the superconductor is much greater than $\xi$ \cite{Belzig}. This condition is nearly fulfilled in $R2_{PtAu}$ ($\xi_{2}  \approx L_{2}/3$). Using the high temperature resistance values  and assuming a gap equal to that of $R2_{PtAu}$ we find $\xi_{1} \approx L_{1}/2$ , $\xi_{4} \approx L_{4}/2$,  $\xi_{3}  \approx 2L_{3}$, and $\xi_{6}  \approx 2L_{6}$. These values explain qualitatively the  reduced transition temperature of $R1_{PtAu}$  and $R4_{PtAu}$ and the absence of a transition for  $R3_{PtAu}$ and  $R6_{PtAu}$.   Moreover we can argue that the superconducting transitions we see are not due to a hidden proximity effect : if the $Al_{2}O_{3}/Pt/Au$ contacts were made superconducting by the laser pulse, the shortest ropes (3 and 6) would become superconducting at temperatures higher than the longer ones (1, 2, 4). It is however not possible to explain the behavior of the sample 5 with the same kind of argument, since the same expression yields a coherence length $\xi_5$ much shorter than the length of the sample, and nonetheless no complete transition is seen. We believe that this is due to the strong disorder in this sample which is very close to the localization limit. Fig.\ref{figurerth} shows that the transition even disappeared after thermal cycling when an increase of room temperature resistance led to complete localization at low temperature. 

\subsection{Role of the number of tubes.}

\begin{figure}[h]

\begin{center}
\includegraphics[width=8cm]{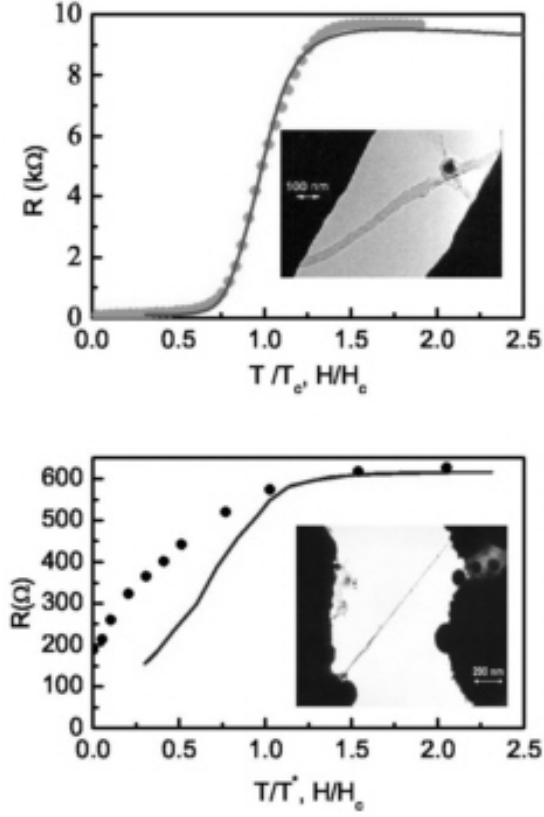}
\end{center}
\caption{
\label{Pt24}
Resistance as a function of temperature (continuous line) and magnetic field (scatter points) for  samples $R2_{PtAu}$ and  $R4_{PtAu}$.
Insets: TEM micrographs of the samples.} 

\end{figure}

Another  a priori  important parameter is the number of tubes in a rope: the less tubes in a rope, the closer the system is to the strictly 1D limit, and the weaker the transition. If we compare the two ropes in Fig. \ref{Pt24}, it is clear that the transition  both in temperature and magnetic field is much broader in the rope $R4_{PtAu}$  with only 40 tubes  than in the rope $R2_{PtAu}$ with 350 tubes. Moreover  there is no inflexion point in the temperature dependence of the resistance in the thinner rope, typical of a strictly 1D behavior. We also expect a stronger screening of e-e interactions in a thick rope compared to a thin one, which could also favor superconductivity as we will discuss below.

\subsection{Role of disorder and intertube coupling}

\begin{figure}[h]

%\begin{center}
\includegraphics[clip=true,width=8cm]{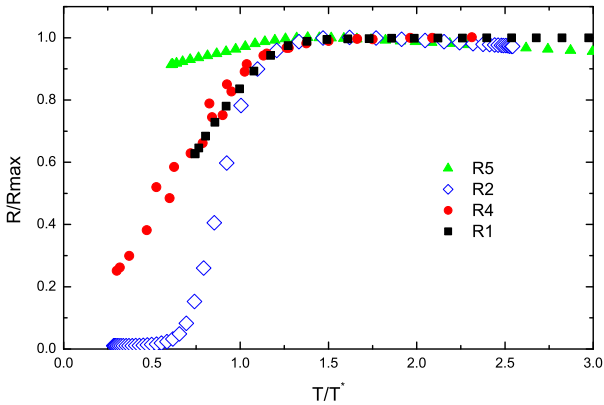}
%\leavevmode
%\epsfxsize=8cm
%\epsfbox{fig1.eps}
%\end{center}
%\vspace{0 cm}
\caption{
\label{scalert}
Resistance as a function of temperature for all samples which undergo a transition, plotted in reduced units $R/R_{max}$ and $T/T^*$.} 

\end{figure}

As is clear from expression (\ref{XI}) for the superconducting coherence length, disorder is at the origin of a reduction of the superconducting coherence length  in a diffusive sample compared to a ballistic one and can in this way also  decrease the destructive influence of the normal contacts. More subtle  and specific to the physics of  ropes,  we have seen that disorder also enhances the intertube coupling, so it can increase the dimensionality of the superconducting transition: weakly disordered ropes like R1 and R4 will be more 1D -like than the more disordered rope R2.  Of course disorder must always be sufficiently small so as not to induce localization. These considerations may explain the variety of behaviors observed and depicted in Fig.\ref{scalert}.

 Finally, disorder is also the essential ingredient which reveals the  difference between the normal state and the superconducting state. In a ballistic rope we would not expect to observe a variation of the resistance over the superconducting transition because in both cases the resistance of the rope is just the contact resistance. 
 
To gain insight in the transport regime, we have performed shot noise measurements of the ropes $R1,3,4,6_{PtAu}$, in the normal state (higher level of 1/f  noise in the more resistive ropes $R2,5_{PtAu}$ made the analysis of shot noise impossible in those samples), between 1 and 15 K. We found a surprisingly strong reduction of the shot noise, by more than a factor 100,  which is  still not well understood, but would indicate  that all the tubes in these ropes are either completely ballistic or completely localized \cite{roche}, in strong apparent contradiction with the observation of the superconducting  transition of $R4_{PtAu}$, with a $60\%$ resistance drop.

It would however be possible to explain a resistance decrease in ballistic ropes turning superconducting, if the number of conducting channels is larger for Cooper pairs than for individual electrons.  In his recent theoretical  investigation of superconductivity in ropes of SWNT, Gonzalez \cite{ gonz02} has shown the existence of a finite  intertube transfer  for   Cooper pairs, even between two tubes of different helicities which have no possibility of single electron intertube transfer. This Cooper pair delocalization could lead to the opening of new channels  when a rope  containing  a mixture of insulating and ballistic tubes becomes superconducting.
%\linebreak
\begin{figure}
\begin{center}
%\leavevmode
%\epsfxsize=6 cm
%\epsfbox{fig4.eps}
\vspace{.8 cm}
\includegraphics[width=9cm]{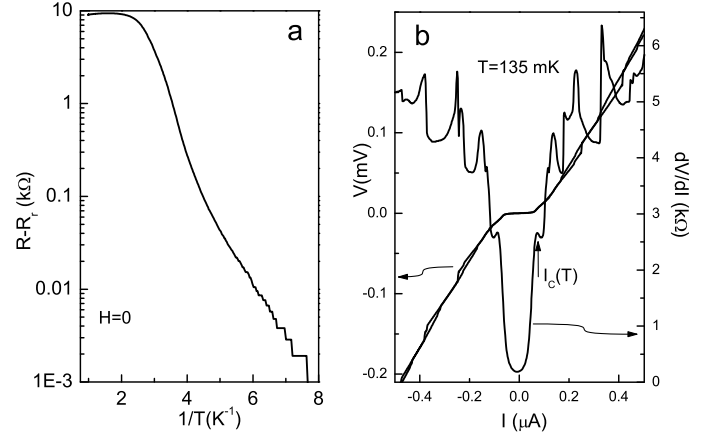}
\end{center}
\caption{
\label{pslips}
a Resistance of $R2_{PtAu}$  plotted on a log scale as a function of the inverse temperature at H=0. We have subtracted the low temperature residual resistance (contact resistance). The slope yields an approximate activation energy of 0.8 K which has to be compared to the condensation energy   $E_{\xi}$ of Cooper pairs  in  a  sample of length $\xi$  $E_{\xi}=M n(E_f) \Delta^2 \xi \simeq 1K$. Note that this characteristic energy is much smaller in our samples containing only one hundred channels than in whiskers where phase slips can only be observed very close to $T_c$.
b- V(I) and $\frac{dV}{dI}(I)$ curves showing the hysteretic behavior in V(I) at each peak in the $\frac{dV}{dI}(I)$  curve for sample $R2_{PtAu}$}
\end{figure}

\subsection{Signatures of 1D superconductivity.} 

Reminiscent of measurements of narrow superconducting metal wires \cite{Giordano}, we find jumps in the differential resistance as the current is increased (Fig. \ref{pslips}). For sample 2 the differential resistance at low currents remains equal to $R_{r}$ up to 50 nA, where it strongly rises but does not recover its normal state value until 2.5 $\mu$A. 
The jump in resistance at the first step corresponds approximately to the normal state resistance of a length $\xi_{2}$ of sample 2. Each peak corresponds to a hysteretic feature in the $V-I$ curve (Fig.  \ref{pslips}b).   These  jumps are identified as phase slips \cite{Giordano,meyer,Tinkham}, which are the occurrence of normal regions located around defects in the sample. Such phase slips can be thermally activated (TAPS), leading to a roughly exponential decrease of the resistance instead of a sharp transition, in qualitative agreement with our experimental observation (Fig. \ref{pslips}a). At sufficiently low temperature, TAPS are expected to be replaced by quantum phase slips, which, when tunneling through the sample, contribute an additional resistance to the zero temperature resistance. 

Also, in sample 2 the current at which the first resistance jump occurs (60 nA, see Fig. \ref{figuredivh}) is close to the theoretical critical current of a diffusive superconducting wire \cite{klap} with gap $\Delta _{2}=85 \mu eV$ ($I_{C}=\Delta_{2} /R_{\xi}e \approx$20 nA), whereas the current at which the last resistance jump occurs (2.4 $ \mu$A, see Fig. \ref{figuredivt}) is close to the theoretical critical current of a ballistic superconducting wire with the same number of conducting channels $I_{C}$$^*=\Delta_{2} /R_{r}e \approx 1$ $ \mu$A  \cite{Tinkham}. 

We also expect this  current $I_{C}$$^*=\Delta_{2} /R_{r}e$ to be the critical current of a structure with this same wire placed between superconducting contacts with gap $\Delta_{S}$ even if $\Delta_{S}< \Delta_{2}$),
and can thus be much larger than the Ambegaokar-Baratoff 
prediction $R_{N} I_{C}\approx \Delta_{S}/e$ \cite{Amb}. Intrinsic superconductivity might thus explain the anomalously large supercurrent measured in the experiments described in the previous section, where nanotubes are connected to superconducting contacts. It is also interesting to compare the low temperature quadratic dependence of the critical currents which is very similar in all samples both on normal and superconducting contacts.

\subsection{Effect of magnetic field} 

It is difficult to say  a priori what causes the disappearance of superconductivity in carbon nanotubes. The value of $H_c(0)$ should be compared to the depairing field in a confined geometry \cite{Meservey}, and corresponds to a flux quantum $\Phi_{0}$ through a length $\xi$ of an individual SWNT of diameter \textit{d}, $\mu_{0}H_{C}=  \Phi_{0}/(2\sqrt{\pi}d\xi)=1.35$ T. But $H_{C}(0) $ is also close to the field $\mu_{0}H_{p}=\Delta /\mu_{B}=1.43$ T at which a paramagnetic state becomes more favorable than the superconducting state \cite{Clogston,Chandrasekhar}. Note that this value is of the same order as the critical field that was measured on SWNT connected between superconducting contacts, i.e. much higher than the critical field of the contacts.

The  linear  dependence of the critical current with magnetic field observed in all samples  is very similar to the data presented in the previous section on SWNT on  superconducting contacts and  appears strongly related to the linear dependence in $T_c(H)$. This linear scaling with magnetic field is surprising since it is not expected to take place in a 1D system and is more typical of 2D superconductivity \cite{smith}.  A depairing mechanism  based on  spin  splitting of the quasiparticle energy states could  however provide a possible explanation. Experiments  performed with various field directions compared to the tube  are thus necessary for a better understanding of the influence of magnetic field on superconductivity in carbon nanotubes.

\section{Conclusion}

Data depicted in the previous section show the existence of  intrinsic superconductivity in ropes of carbon nanotubes which number of tubes varies between 30 and 400. The question of the existence of superconducting correlations  in the  limit of the individual tube cannot be answered yet.  It is of course tempting to consider the high supercurrent measured  on superconducting contacts as a strong indication that superconducting fluctuations are present also in individual carbon nanotubes. However  since unscreened Coulomb repulsive interactions in these samples are expected to suppress superconductivity,   precise investigations of  individual carbon nanotubes on normal contacts are necessary. It is essential to conduct experiments on sufficiently long samples (like the ropes presently studied)  so that intrinsic superconductivity is not destroyed by the normal contacts. Note that recent magnetization experiments  \cite{tang} also strongly support the existence of superconducting fluctuations below 10 K in  very small diameter (0.4 nm) individual tubes grown in zeolites.

We now discuss what could be the relevant mechanism for superconductivity in carbon nanotubes. Observation of
superconductivity in carbon based compounds was reported a long time ago. First in graphite intercalated with alkali atoms (Cs,K),  superconducting transitions were observed  between 0.2 and 0.5 K \cite{Hannay}. Much higher temperatures were observed in alkali-doped fullerenes \cite{Gunnarsson} because of the coupling to higher energy phonons.  In  all these experiments it was essential to chemically dope the system to observe superconductivity. There is no such chemical dopants in the ropes of carbon nanotubes studied here. As shown in previous works there is some possibility of  hole doping of the tubes by the gold metallic contacts which electronic work function is 
larger than in the tubes  \cite{Venema,Odinstov}.  However, although this doping could slightly depopulate the highest occupied energy band in a semiconducting tube it is very unlikely that it is strong enough to depopulate other  lower energy subbands for a metallic tube with a diameter in the nm range.  
More interesting would be a mechanism related  to the 1D electronic structure of carbon nanotubes. A purely electronic coupling mechanism has been indeed shown to induce superconducting fluctuations in coupled double chain systems such as ladders \cite{Schultz}. The relevance of this mechanism has been considered also in  carbon nanotubes away from half filling but the very small order of magnitude for the energy scale of these superconducting fluctuations is not compatible with our findings \cite{Egger}.

Recent estimations of the electron phonon coupling coupling constants \cite{gonz02}, \cite{bourbonnais}, \cite {egger2} in carbon nanotubes seem to be more promising. It is shown that the breathing modes specific to carbon nanotubes can be at the origin of a strong electron phonon coupling giving rise to attractive interactions which can possibly overcome repulsive interactions in very small diameter tubes .   The possible coupling of these rather high energy modes  to  low energy compression modes in the nanotube have been also considered, following the Wenzel Bardeen singularity scheme  initially propose by Loss and Martin where   low energy phonons are shown to  turn repulsive interactions in a Luttinger liquid into attractive ones and drive the system towards a superconducting phase \cite{Loss}. The suspended character of the  samples may be essential in this mechanism.

A.K. thanks the Russian foundation for basic research and solid state nanostructures for financial support, and thanks CNRS for a visitor's position. We thank M. Devoret, N. Dupuis, T. Giamarchi, J. Gonzalez, T. Martin, D. Maslov, C. Pasquier for stimulating discussions, and P. Lafarge and L. L\'evy for the high field measurements at LCMI.

\vspace{-.5cm}

\begin{appendix}
\section{Proximity induced superconductivity in multiwall carbon nanotubes.}
\begin{figure}[h]
\includegraphics[clip=true,width=8cm]{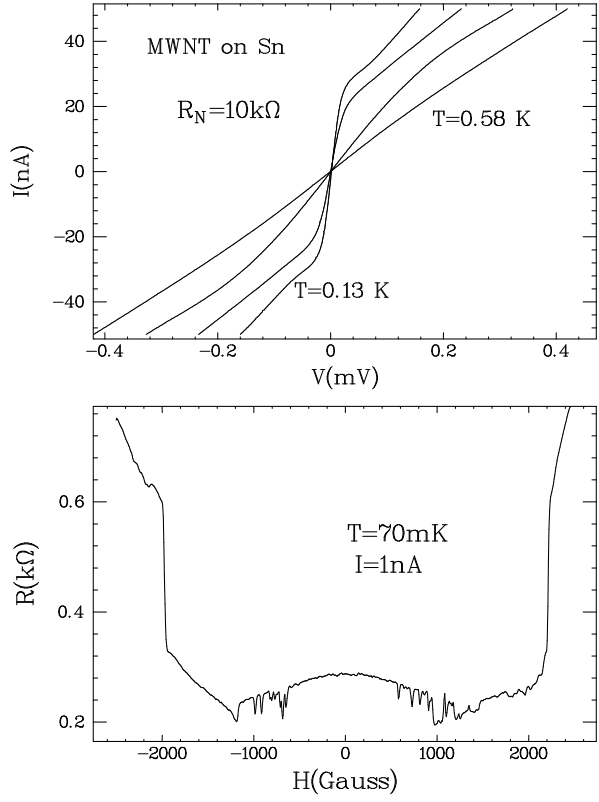}

%\[	\epsfbox{file.eps}	\]
		\caption{Top: I V curves  of a multiwall nanotube  mounted on tin contacts  for various
values of temperature. The characterisation  by  transmission electron microscopy yields the diameters of the external and smallest internal shell respectively equal to: 25 and 12 nm  which corresponds to a multiwall nanotube of approximatively 40 shells. Bottom: Magnetoresistance measured at 60mK.  \label{ivmw}}
		\end{figure}

We have also investigated multiwall carbon nanotubes on tin contacts. We show in Fig. \ref{ivmw} the  IV curves recorded at several temperatures on a multiwall carbon nanotube with a normal state resistance of $10~k~\Omega$. The curves do not exhibit a clear Josephson behavior like in ropes but only  linearities in the IV curves. It is  however possible to define a characteristic current of the order of 30 nA  ( i.e. very small compared to the values of critical currents observed in ropes). Moreover there is no sign of phase slips.
The magnetoresistance is a non monotonous function of magnetic field, this behavior has already been observed in amorphous superconducting wires. 

\end{appendix}
\end{document}